\documentclass[a4paper,11pt]{article}
\usepackage{pos}
\usepackage{enumitem}
\usepackage{wrapfig}
\usepackage[most]{tcolorbox}
\usepackage{caption}
\usepackage{graphicx}
\usepackage{subcaption}
\usepackage{comment}

\title{What's new in NuRadioMC: Multilayer Analytic Raytracer}
\author*[a,b]{H.~Warnhofer}
\author[b]{S.~Bouma}
\author[c]{V.~de Henau}
\author[d,e]{B.~Hendricks}
\author[a,b]{J.~Henrichs}
\author[f]{N.~Heyer}
\author[a]{E.~Huesca~Santiago}
\author[a,b]{A.~Jaitly}
\author[b]{P.~Laub}
\author[a,b]{A.~Nelles}
\author[f]{M.~Ravn}
\author[g]{F.~Schl{\"u}ter}
\author[a,b]{Z.S.~Selcuk}
\author[b]{K.~Terveer}
\author[f,h]{C.~Glaser}
\author[d,e,i]{C.~Welling}
\author[a,j]{P.~Windischhofer}

\affiliation[a]{Deutsches Elektronen-Synchrotron DESY, Platanenallee 6, 15738 Zeuthen, Germany}

\affiliation[b]{Erlangen Centre for Astroparticle Physics (ECAP), Friedrich-Alexander-University Erlangen-N\"urnberg, \\91058 Erlangen, Germany}

\affiliation[c]{Vrije Universiteit Brussel, Dienst ELEM, B-1050 Brussels, Belgium}

\affiliation[d]{Center for Multimessenger Astrophysics, Inst.\ of Gravitation and the Cosmos, Pennsylvania State University, University Park, PA 16802, USA}

\affiliation[e]{Dept.\ of Physics, Pennsylvania State University, University Park, PA 16802, USA}

\affiliation[f]{Uppsala University, Dept.\ of Physics and Astronomy, Uppsala, SE-752 37, Sweden}

\affiliation[g]{Universit\'e Libre de Bruxelles, Science Faculty CP230, B-1050 Brussels, Belgium}

\affiliation[h]{Dept.\ of Physics, TU Dortmund University, Dortmund, Germany}

\affiliation[i]{Dept.\ of Astronomy and Astrophysics, Pennsylvania State University, University Park, PA 16802, USA}

\affiliation[j]{Dept.\ of Physics, Dept.\ of Astronomy \& Astrophysics, Enrico Fermi Inst., Kavli Inst.\ for Cosmological Physics, University of Chicago, Chicago, IL 60637, USA}

\emailAdd{hannes.warnhofer@desy.de}
\abstract{NuRadioMC is a framework for the simulation of ultra-high-energy neutrino detectors that measure the radio signal emitted in neutrino-induced particle cascades used in different radio neutrino experiments. We present an extension to the analytic raytracing method for multilayered exponential refractive index models, where the full refractive index profile is described by a set of different single-exponential layers, supporting realistic medium descriptions at various experiment sites while maintaining computational efficiency. This approach also enables a realistic exponential refractive index description of the atmosphere and allows us to model signal propagation over non-smooth changes in the refractive index. This report outlines the fundamentals of the multilayered analytic raytracing method and shows some applied examples.}
\FullConference{11th International Workshop on Acoustic and Radio EeV Neutrino Detection Activities (ARENA2026)\\
8-11 June 2026\\
Karlsruhe, Germany\\}

\begin{document}
\maketitle
%\section{Introduction}
%\include{introduction}

\section{Introduction to NuRadio}
NuRadio is a Python-based open-source simulation and reconstruction framework for radio detectors of ultra-high-energy neutrinos or cosmic rays. It provides functionality to simulate the expected development and propagation of a radio signal following a neutrino or cosmic ray interaction in ice, to calculate the detector response to such a signal, to simulate internal and external thermal noise in the detector, as well as different tools for reconstructing the vertex position, direction, and energy of the incoming particle. 

The software for simulating the radio signal as it arrives at the detector is contained in the \textbf{\texttt{NuRadioMC}} package. Aiming at high flexibility, the software is organized in a modularized manner, with each module containing functionality dedicated to a certain part of the simulation chain: from modeling neutrino arrival fluxes and the production of secondaries (\texttt{NuRadioMC.EvtGen}) over modeling the radio signal expected from the Askaryan effect (\texttt{NuRadioMC.SignalGen}) to simulating the propagation of the radio signal through ice and air (\texttt{NuRadioMC.SignalProp}). Within each module, different methods are available to be used in the corresponding part of the simulation chain, for example, different flux models, various models for the generation of the radio signal, and different methods of simulating the signal propagation. Furthermore, the package provides a number of refractive index and attenuation models for various experiment sites. For a more detailed overview, see \cite{glaserNuRadioMCSimulatingRadio2020a}. 

The \textbf{\texttt{NuRadioReco}} package contains all the detector response descriptions needed to calculate the expected voltage trace in the detector circuit arising from a present electromagnetic field. It allows us to simulate the filtering and triggering of the detector setup, providing a full description of the expected detection behavior. Furthermore, the package holds the reconstruction algorithms that enable different methods of reconstructing the interaction vertex position, the direction, and the initial energy of a neutrino or cosmic ray interaction. For deeper insight into the reconstruction principles and the general data structure of \texttt{NuRadio}, we advise reading \cite{glaserNuRadioRecoReconstructionFramework2019}. 

The software was developed to provide a common framework for in-ice neutrino detectors such as ARA, ARIANNA and RNO-G. However, the generalized software architecture is not limited to certain experiments, but enables one to analyze various different detector setups and to investigate different parts in the simulation process, highlighting \texttt{NuRadio}'s flexible capabilities for generic detector studies. Building on experience with RNO-G, \texttt{NuRadio} is also planned to provide the analysis framework for the radio component of IceCube-Gen2. Since 2025, \texttt{NuRadioReco} also includes the descriptions of the air-shower cosmic-ray detectors LOFAR and SKA and is incorporated in their corresponding analysis pipelines \cite{Bouma:2025cM}.
\begin{tcolorbox}[
    enhanced,
    width=\linewidth,
    colback=gray!10,
    colframe=black,
    arc=3mm,
    boxrule=0.6pt,
    left=2mm,
    right=2mm,
    top=1mm,
    bottom=1mm,
    halign=center
]
\begin{itemize}[nosep,leftmargin=*]
    \item[] \textbf{\texttt{pip install nuradiomc}}
    \item[] GitHub: \textbf{\texttt{https://github.com/nu-radio/NuRadioMC}}
    \item[] Documentation: \textbf{\texttt{https://nu-radio.github.io/NuRadioMC/main.html }}
\end{itemize}
\end{tcolorbox}
\section{Signal Propagation in NuRadioMC}
As is the case for most particle physics experiments, the radio detection of ultra-high-energy neutrinos is based on Monte Carlo simulations, which require the detailed description of all elements in the signal chain: from the initial neutrino interaction all the way to the detector response. An important part is to precisely model the path that the radio signal takes after it is induced from a particle shower. To simulate the evolution of the signal, the first step is to find the physical path that an electromagnetic signal is expected to take. When dealing with long-range radio signals, we can use the large-wavelength approximation of geometric optics, allowing us to describe the evolution of the electromagnetic wave by a single ray, which is representative for the propagating wavefront.  

\begin{wrapfigure}[14]{r}{0.48\textwidth}
    \centering
    \includegraphics[width=\linewidth]{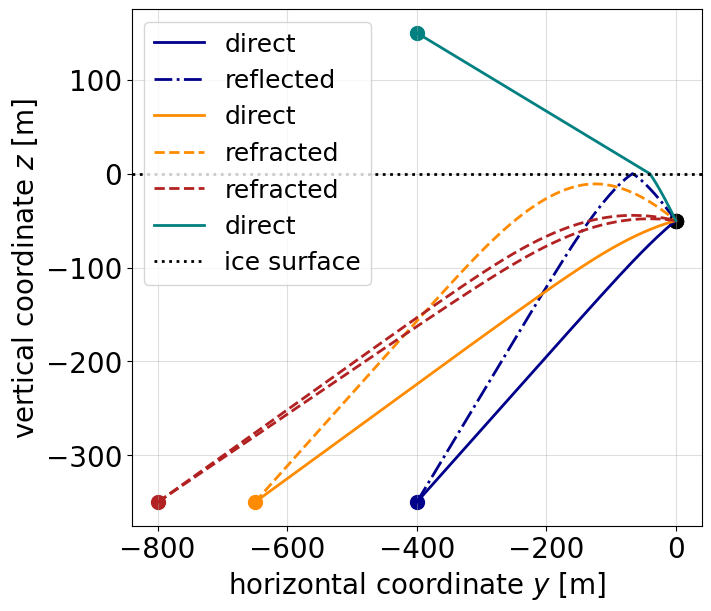}
    \caption{Examples of the different ray path combinations that can occur.}
    \label{fig:examplerays}
\end{wrapfigure}

In a medium with changing refractive index, like glacial ice, upward-going signal paths will bend in shallower depths and turn back down again. In NuRadioMC, paths that have a turning point before reaching the receiver are referred to as \textit{refracted}, if not we call them \textit{direct}. We generally expect two path solutions for an in-ice signal, which can either be a direct ray and a refracted ray, two refracted rays or a direct ray and one that gets reflected at the ice surface. Note, however, that depending on the refractive index model, there are regions where more than two path solutions can be found. Figure \ref{fig:examplerays} shows examples for the classically expected path solutions.

\setlength{\intextsep}{2pt} 

For both path solutions, we need to evaluate the following path specific parameters:
\begin{itemize}[noitemsep,topsep=2pt,parsep=1pt,partopsep=0pt]
    \item Signal travel time
    \item Viewing angle at the interaction
    \item Arrival angle at the antenna
    \item Frequency-dependent attenuation
    \item Possible reflection loss at the surface
    \item Focusing factor
\end{itemize}
All of these parameters are needed for a realistic description of the expected signal at the detector and provide crucial information for reconstruction algorithms and almost all types of analyses. For the successful detection of neutrino interactions, it is therefore crucial to have a well-understood and accurate simulation of the signal propagation.

The \texttt{NuRadioMC.SignalProp} module includes different methods for this task, which differ in terms of possible complexity of the medium and computation speed. On one hand, we can use the \texttt{radiopropa} raytracer which is based on \texttt{RadioPropa}, a numerical solver that iteratively solves coupled differential equations step by step to find the physical path that minimizes travel time from emitter to receiver. While this allows for a high flexibility regarding medium definitions, including variable discontinuity layers and reflections, the computation takes comparably long, reaching up to \textbf{$2\,\mathrm{s}$} for long ray paths, making it computationally infeasible for large-scale Monte-Carlo simulations. 

On the other hand, we have the \texttt{analytic} raytracer, which assumes a simple exponential refractive index model (see next section) for which an analytical solution of the path can be found. This enables a much quicker solution finding and parameter calculation, taking only around \textbf{$0.15\,\mathrm{ms}$}. However, a simple exponential refractive index model seems to be insufficient to model the measured situations at real experiment sites, which can be seen in figure \ref{fig:refractiveindex} \cite{windischhoferCalibratingRadioNeutrino2024b}. 

To be able to include a more realistic ice model in simulations and reconstructions, an expansion of the \texttt{analytic} raytracer was recently developed, which uses the same solution finding algorithms but allows for medium descriptions that consist of multiple layers of exponential models. The solution finding of the \texttt{multilayer} raytracer using a 3-layer-model takes around \textbf{$0.4\,\mathrm{ms}$}, which is around 3 times slower than the single layer case but still sufficiently fast for use in large MC productions. The computation time scales roughly with the number of layers. In the following section, we will have a detailed look into the principles behind the analytic raytracing approach in general and the multilayer case in particular.

\section{Multilayer Analytic Raytracer}
This section aims to provide a general overview about the working principles of the multilayer analytic raytracer, which is an expansion of the single layer analytic raytracer. The following derivation is in large parts based on the corresponding descriptions in appendix C of \cite{glaserNuRadioMCSimulatingRadio2020a}, which contains a more detailed discussion of the underlying theory.

The propagation of light in a dielectric medium follows Fermat's principle, which states that light takes the path of least duration. Mathematically, we can write it as follows:
\begin{equation}
     S=\int_A^Bn\,ds\,\,\,\,\,\,\,\mathrm{and}\,\,\,\,\,\,\,\delta S =0
\end{equation}
Assuming that the refractive index only depends on the depth, i.e. $n=n(z)$, we can rotate the coordinate system so that $\partial_z x =dx/dz=0$. In practice, this allows us to reduce every 3D raytracing problem to a 2D problem in the vertical plane that contains the start and end point. Mathematically, we can rewrite Fermat's principle using $ds=\sqrt{dx^2+dy^2+dz^2}=dz\,\sqrt{(\partial _z y)^2+1}$:
\begin{align}
    \delta \int^B_A n(z)\,\sqrt{1+(\partial_z y)^2}\,dz=0
\end{align}
By making a variable change, we can rearrange and integrate:  
\begin{align}\label{eq:de}
    \ln{\partial_z y}-\frac{1}{2}\,\ln{((\partial_z y)^2+1)}=-\ln{n}+constant \\
    \Rightarrow \frac{\partial_z y}{\sqrt{(\partial_z y)^2+1}}\,n(z)=\frac{1}{C_0}=constant
\end{align}
In general, this is the equation that path solutions need to satisfy. Note how this already shows that there is a physical invariant in this problem, which appears as our integration constant $C_0$. We can think of the local $\partial_z y$ as the angle between the ray at a given point with respect to the horizontal ($\tan{(\partial_z y)(z)}=\theta(z)$), which allows us to rewrite the differential equation from above as a local Snell's law, using trigonometric relations:
\begin{equation}\label{eq:angle}
    \sin{(\theta(z))}\cdot n(z)=\frac{1}{C_0}=constant
\end{equation}
While the \texttt{radiopropa} raytracer solves these equations numerically along the path, step by step, the analytic approach makes use of a special form of refractive index models, which allows us to take advantage of the exponential function's special analytical properties. This is what is referred to as an \textbf{exponential refractive index model} in the following:
\begin{equation}\label{eq:exp}
    n(z)=n_\mathrm{ice}-\Delta_n\exp{(z/z_0)}
\end{equation}
For this $n(z)$, an analytical solution for the horizontal displacement $y$ as a function of depth $z$ that satisfies equation \ref{eq:de} can be found. With $\gamma=\Delta_n\exp{(z/z_0)}$:  
\begin{equation}
    y(z)=\pm C_0^{-1}c^{-1/2}z_0\ln{(F(\gamma))}\mp z_0C_1'
\end{equation}
\begin{equation}
    F(\gamma)=\frac{\gamma}{2\sqrt{c\left(\gamma^2-2\gamma n_\mathrm{ice}+c\right)}-2\gamma n_\mathrm{ice}+2c}
\end{equation}
For our task, it can help to think of this solution in the way that the displacement can be found from:
\begin{equation}\label{eq:y}
    y(z)=Y(z,C_0,n_\mathrm{ice},\Delta_n,z_0)+C_1(C_0,\vec{x}_\mathrm{start}) 
\end{equation}
In general, the ray path for a given refractive index model is described by the shape parameter $C_0$ and the horizontal offset $C_1$, which depends on the coordinates of the starting point. The function $y(z)$ has a singularity where the ray becomes horizontal (when $C_0\cdot n(z)=1$) and has a nonphysical solution branch above. However, after becoming horizontal and turning downward, the path is symmetrical to the upward propagation, which is why we can simply mirror the path vertically at the turning point to describe the signal propagation after turning. 
\begin{figure}
    \centering

    \begin{subfigure}{0.42\textwidth}
        \centering
        \includegraphics[width=\linewidth]{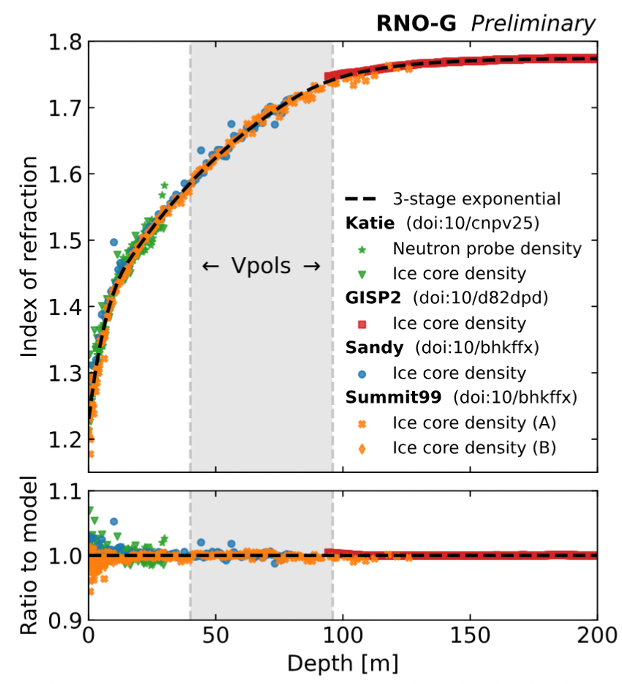}
        \label{fig:first}
    \end{subfigure}
    \hfill
    \begin{subfigure}{0.52\textwidth}
        \centering
        \includegraphics[width=\linewidth]{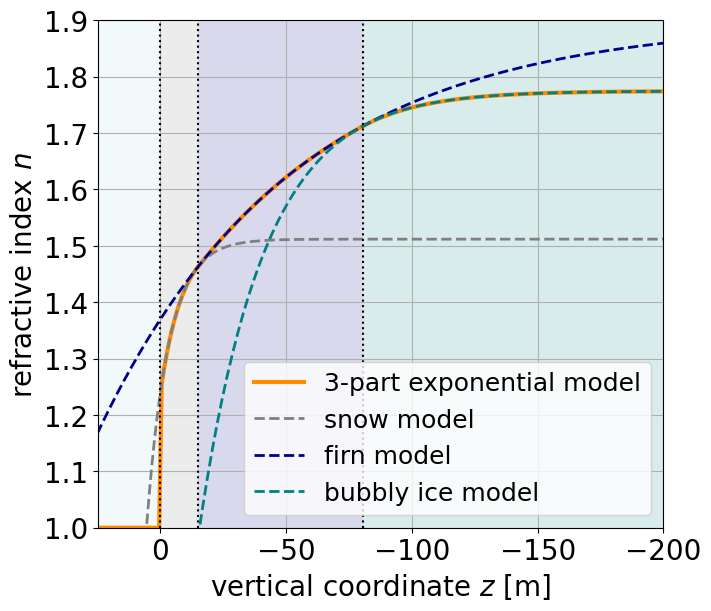}
        \label{fig:second}
    \end{subfigure}

    \caption{Measurements of the refractive index at Summit Station together with the 3-layer exponential refractive index model \textit{(left)}. Plot taken from \cite{windischhoferCalibratingRadioNeutrino2024b}. Detailed look at the 3-layer model including the individual constituents \textit{(right)}.}
    \label{fig:refractiveindex}
\end{figure}

In terms of raytracing, the goal is to quickly and precisely find the two path solutions for a start point $(y_1,z_1)$ and end point $(y_2,z_2)$ in a given medium. We can evaluate the analytical solution $y(z)$ at the receiver depth $z_2$ and minimize the difference $\Delta y=y(z_2)-y_2$ using numerical minimization algorithms to find the corresponding $C_0$ that defines the signal path that connects the two points. The solution finding is done by first using \texttt{scipy.optimize.root} to minimize $(\Delta y)^2$ and then looking for higher $C_0$ solutions by minimizing $\Delta y$ with \texttt{scipy.optimize.brentq} algorithm. This approach was already developed for the single layer analytic raytracer to ensure a stable solution finding and could be reused in a slightly adapted way for the multilayer case.

We can expand the refractive index model to multiple layers of exponential models, where the parameters differ between the layers, but $n(z)$ is still described according to equation \ref{eq:exp} within each layer. This allows us to describe the ray path within each layer analytically and to iteratively evaluate the entire path layer by layer by calculating the horizontal offset parameters $C_1$ for each layer, ensuring continuity in the ray path. The displacement $y(z)$ can still be described as in equation \ref{eq:y} but with different $n_\mathrm{ice}$, $\Delta_n$, $z_0$ and $C_1$ for each layer. We are now able to evaluate signal paths for all kinds of refractive index models that are described by a series of exponential layers, like the 3-layer model shown in figure \ref{fig:refractiveindex}. The refractive index must increase monotonically with lower $z$, but it does not have to be continuous, Snell's law is naturally encoded. In practice, this also simplifies the calculation of ice-to-air ray paths and allows for a more detailed description of the atmosphere. However, note that the solution finding algorithm is optimized for refractive index models that resemble natural glacial ice like in Greenland or at the South Pole and is not stable for too large jumps of $n$ at discontinuities within the ice. Furthermore, only reflection at the ice surface at $z=0$ is implemented in the current state, not at possible discontinuities within the ice.
\begin{figure}
    \centering
    \includegraphics[width=\linewidth]{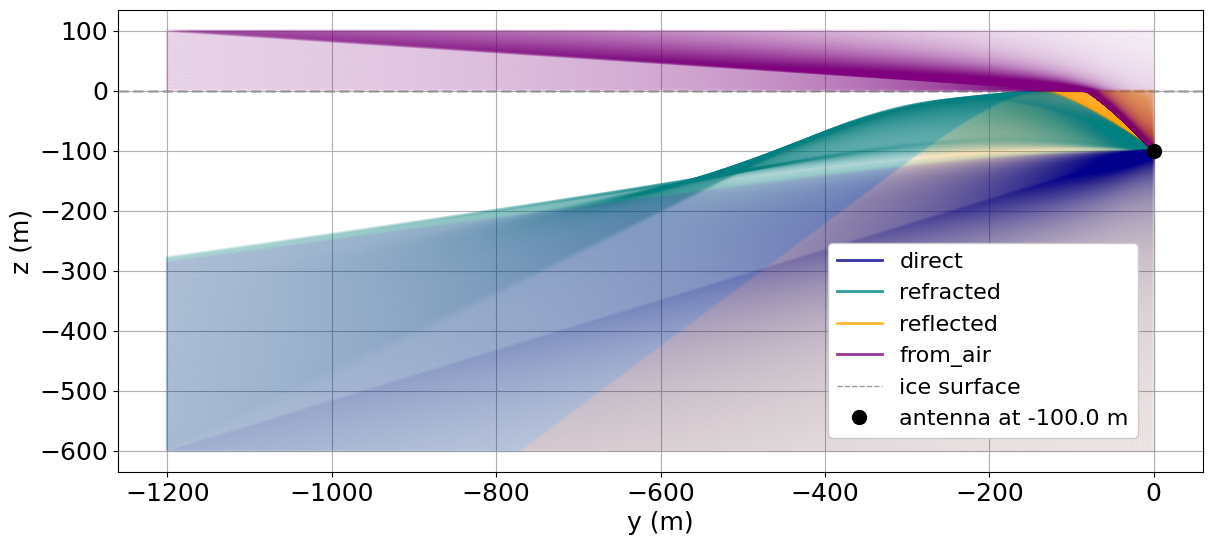}
    \caption{Ray paths from 1500 starting points located along the lower ($z=-600\,\mathrm{m}$), left ($y=-1200\,\mathrm{m}$), and upper ($z=+100\,\mathrm{m}$) side (500 each) towards the antenna at $(y=0\,\mathrm{m},z=-100\,\mathrm{m})$, using the \texttt{greenland\_3exp} ice model. Color coded regarding whether the rays are direct, refracted, reflected at the surface or coming from the air, passing through the surface.}
    \label{fig:niceplot}
\end{figure}

After finding the $C_0$s for the correct path solutions, the calculation of the corresponding path lengths, signal travel times, or focusing factor can be reduced to a piecewise evaluation of the integrals described in \cite{glaserNuRadioMCSimulatingRadio2020a} and the appendix of \cite{boumaDirectionReconstructionRadio2025}. The launch angle, the reflection angle, and the receiving angle can be calculated directly from equation \ref{eq:angle}. Using \texttt{numba} support, solution finding can be done in around $0.4\,\mathrm{ms}$ and parameter calculation takes around $15\,\mathrm{\mu s}$ for each parameter per solution pair. Only the attenuation has to be calculated numerically and depends on the signal frequency, making it as "slow" as the solution finding. Figure \ref{fig:niceplot} shows the paths for 1500 starting points spread out on the lower, left, and upper side of the plot canvas (500 points each), towards an antenna at $z=-100\,\mathrm{m}$. We can see distinctive features, like the shadow zone, from where no signals are expected within the geometric optics approximation, or the caustic pocket, from where more than two path solutions are expected.

\begin{figure}
    \centering

    \begin{subfigure}{0.48\textwidth}
        \centering
        \includegraphics[width=\linewidth]{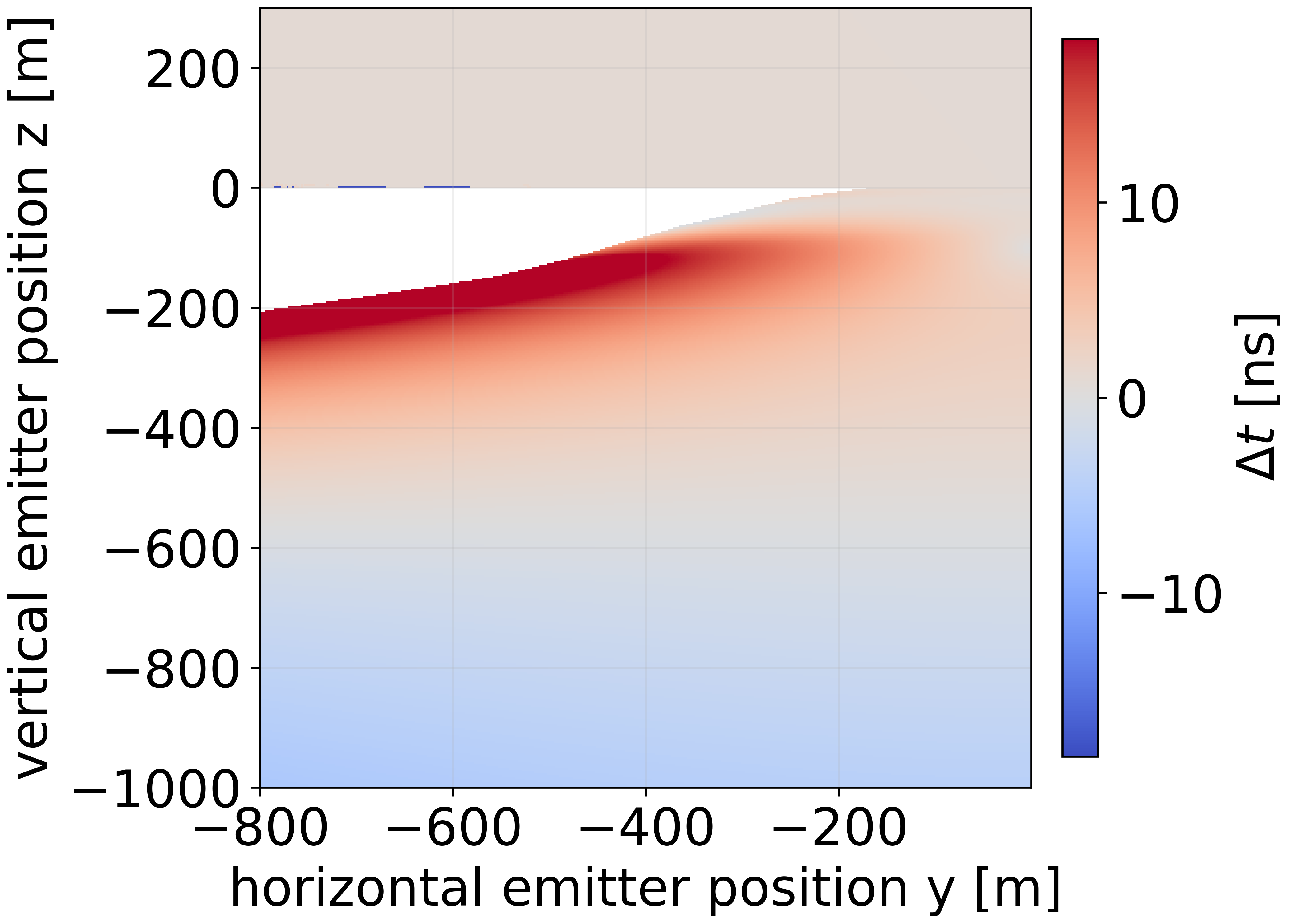}
        \label{fig:time}
    \end{subfigure}
    \hfill
    \begin{subfigure}{0.48\textwidth}
        \centering
        \includegraphics[width=\linewidth]{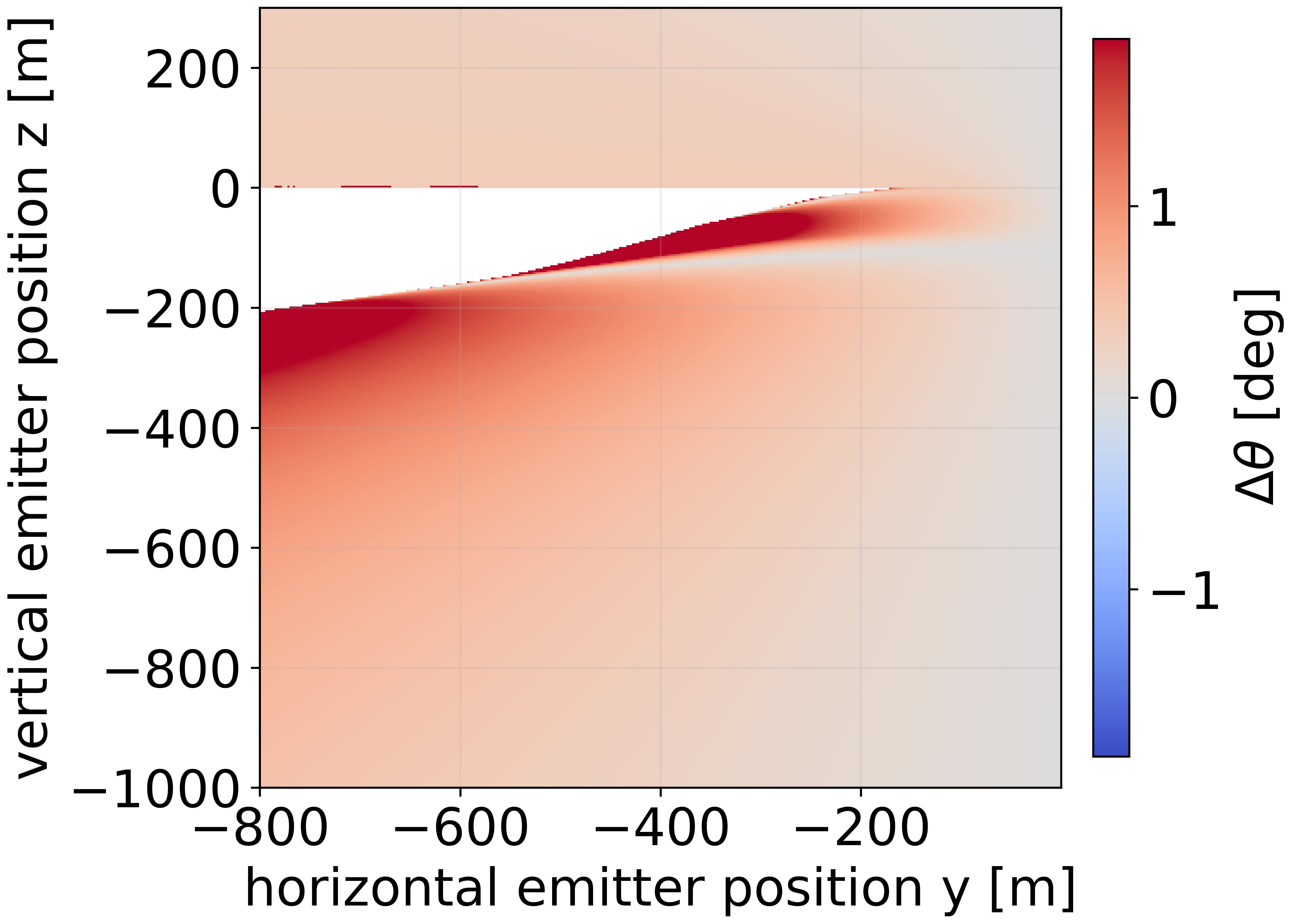}
        \label{fig:angle}
    \end{subfigure}

    \caption{Difference in signal travel time \textit{(left)} and arrival angle at the antenna \textit{(right)} between the \texttt{greenland\_simple} and \texttt{greenland\_3exp} for an antenna at $(y=0\,\mathrm{m},z=-100\,\mathrm{m})$. Results for the first (lower $C_0$) of the two path solutions are compared.}
    \label{fig:timeandanglediff}
\end{figure}

Comparisons with the established version of the single layer \texttt{analytic} raytracer show exact agreement, and the comparison to \texttt{radiopropa} show the same minor deviations that can be observed in the single layer raytracer. We can use this new tool, for example, to compare the differences in travel time that we expect between the single layer \texttt{greenland\_simple} ice model and the now available \texttt{greenland\_3exp} model, using the newest fit parameters presented in \cite{Heyer:2026}, which can be seen in figure \ref{fig:timeandanglediff}.
\section{Summary}
The NuRadio simulation framework continues to be developed further. A new raytracing method was added to the \texttt{NuRadioMC.SignalProp} module, which enables the usage of layered exponential refractive index models and therefore more flexible and realistic simulations and reconstructions of radio signal paths for in-ice radio detectors. The method is implemented as a 2D raytracing class that replaces the "old" single layer analytic raytracer within the existing 3D analytic raytracing class when a medium of type \texttt{IceModelExpLayers} is used. It integrates seamlessly within existing \texttt{NuRadio}-based analyses when switching to a new multilayer medium. Generally, the raytracer is very easy to use, depends only on established Python libraries, and can be directly used after installing NuRadioMC with \texttt{pip install nuradiomc}. New medium models with an arbitrary number of layers (above and beyond the ice surface) can be defined, or one of the provided models for Greenland, Moore's Bay, or the Southpole can be used. The new raytracer allows us to quickly simulate the propagation of radio signals in realistic ice. The computation time scales approximately with the number of layers and takes around $t\approx n_L\cdot 0.15\,\mathrm{ms}$ for each pair of points. 

Besides the new raytracing module, many other important developments have been implemented in NuRadio in the past few years. In addition to many smaller improvements in performance and simulation flexibility, the LOFAR detector response and corresponding functionality have been integrated \cite{Bouma:2025cM}, a new likelihood reconstruction was added \cite{ravn2026likelihoodreconstructionradiodetectors}, the treatment of birefringence in ice models was included, improved noise simulations were added, and an optimized CoREAS read-in was developed, which also includes interpolation for arbitrary antenna positions, allowing for a smoother and more flexible integration of CoREAS simulations. Overall, the NuRadio framework continues to develop, and new features for both simulation and reconstruction are being added, advancing the possibilities it provides for the usage in radio based neutrino and cosmic-ray experiments. We would like to thank the European Union for the support within the European Unions Horizon 2020 research and innovation program (ERC, Pro-RNO-G No 101115122 and NuRadioOpt No 101116890). 

\bibliographystyle{plainnat}
\bibliography{bib}

\end{document}